\documentclass[10pt,ngerman,english]{wlscirep}  
\usepackage[utf8]{inputenc}
\usepackage{amsfonts,amssymb,amsmath}
\usepackage{graphicx}

\def \draft {1}

\usepackage{lineno}

\usepackage{xparse}
\usepackage{ifthen}
\DeclareDocumentCommand{\comment}{m o o o o}
{\ifthenelse{\draft=1}{
    \textcolor{red}{\textbf{C : }#1}
    \IfValueT{#2}{\textcolor{blue}{\textbf{A1 : }#2}}
    \IfValueT{#3}{\textcolor{ForestGreen}{\textbf{A2 : }#3}}
    \IfValueT{#4}{\textcolor{red!50!blue}{\textbf{A3 : }#4}}
    \IfValueT{#5}{\textcolor{Aquamarine}{\textbf{A4 : }#5}}
 }{}
}

\newcommand{\todo}[1]{
\ifthenelse{\draft=1}{\textcolor{red!50!blue}{\textbf{TODO : \textit{#1}}}}{}
}

\usepackage{wrapfig}
\usepackage{floatrow}    
\usepackage{subfig} 

\usepackage{array}
\usepackage{multirow}

\usepackage{adjustbox} 		
\usepackage{upgreek}		
\usepackage{units}		
\usepackage{mathtools}			
\usepackage{xargs}                      

\newcommand{\Si}{\text{Si}}
\newcommand{\Ci}{\text{Ci}}

\newcommand{\etal}{\textit{et al. }}

\usepackage{lipsum}
\usepackage{booktabs}  

\usepackage{accents}      

\title{
 Influence of flicker noise and nonlinearity on the frequency spectrum of spin torque nano-oscillators        }

\author[1,*]{Steffen Wittrock}
\author[1,2]{Philippe Talatchian}
\author[3]{Sumito Tsunegi}
\author[1]{Denis Crété}
\author[3]{Kay Yakushiji}
\author[1]{Paolo Bortolotti}
\author[4]{Ursula Ebels}
\author[3]{Akio Fukushima}
\author[3]{Hitoshi Kubota}
\author[3]{Shinji Yuasa}
\author[1]{Julie Grollier}
\author[5]{Gilles Cibiel}
\author[6]{Serge Galliou}
\author[6]{Enrico Rubiola}
\author[1,$\dagger$]{Vincent Cros}

\affil[1]{Unit\'{e} Mixte de Physique CNRS/Thales, Univ. Paris-Sud, Univ. Paris-Saclay, 1 Avenue Augustin Fresnel, 91767 Palaiseau, France}
\affil[2]{Institute for Research in Electronics and Applied Physics, Univ. of Maryland, College Park, MD, USA   }
\affil[3]{National Institute of Advanced Industrial Science and Technology (AIST), Spintronics Research Center, Tsukuba, Ibaraki 305-8568, Japan}
\affil[4]{Univ. Grenoble Alpes, CEA, INAC-SPINTEC, CNRS, SPINTEC, 38000 Grenoble, France}
\affil[5]{Centre National d'\'{E}tudes Spatiales (CNES), 18 av. Edouard Belin, 31401 Toulouse, France}
\affil[6]{FEMTO-ST Institute, CNRS, Univ. Bourgogne Franche Comt\'{e}, 25030 Besançon, France}

\affil[*]{steffen.wittrock@cnrs-thales.fr}
\affil[$\dagger$]{vincent.cros@cnrs-thales.fr}

\begin{abstract}

The correlation of phase fluctuations in any type of oscillator fundamentally defines its spectral shape. 
However, in nonlinear oscillators, such as spin torque nano oscillators, the frequency spectrum can become particularly complex. 
This is specifically true when not only considering thermal but also colored $1/f$ flicker noise processes, which are crucial in the context of the oscillator's long term stability.
In this study, we address the frequency spectrum of spin torque oscillators in the regime of large-amplitude steady oscillations experimentally and as well theoretically. We particularly take both thermal and flicker noise into account.
We perform a series of measurements of the phase noise and the spectrum on spin torque vortex oscillators, notably varying the measurement time duration. Furthermore, we develop the modelling of thermal and flicker noise in Thiele equation based simulations. 
We also derive the complete phase variance in the framework of the nonlinear auto-oscillator theory and deduce the actual frequency spectrum. We investigate its dependence on the measurement time duration and compare with the experimental results.
Long term stability is important in several of the recent applicative developments of spin torque oscillators. This study brings some insights on how to better address this issue.

\end{abstract}

\begin{document}

\flushbottom
\maketitle

\thispagestyle{empty}


\section{Introduction}

Spin torque nano oscillators (STNOs) are nano-scale devices, which use spin polarized dc currents to drive a steady state electrical rf signal. Thereby, they exploit the spin transfer torque effect 
to excite the magnetization dynamics of a magnetic layer in a nanostructure. Those are converted into a dynamical change of the device resistance mapping the magnetic dynamics through the magnetoresistive effect existing in spintronic devices. 
Key property of STNOs is their nonlinearity, i.e. a coupling between the oscillator's amplitude and phase, which is an intrinsic effect of magnetic dynamics. Therefore, STNOs provide the unique opportunity for studying nonlinear dynamics at the nanoscale\cite{Slavin2009}. Furthermore, and also linked to their nonlinearity, they are considered as promising candidates for next-generation multifunctional spintronic devices\cite{Locatelli2013,Ebels2017}. In addition to their nanometric size ($\sim 100\,$nm), STNOs in general benefit from a high frequency tunability and compatibility with standard CMOS technology\cite{Kreissig2017}. Potential applications are manifold and reach from high data transfer rate hard disk reading\cite{Sato2012} and wide-band high-frequency communication\cite{Muduli2010,Choi2014,Purbawati2016,Ruiz-Calaforra2017,Litvinenko2019} to spin wave generation\cite{Demidov2010,Madami2011} for e.g. magnonic devices\cite{Kruglyak2010,Chumak2017}. 
More recently, STNOs have also been identified as key elements in the realization of broadband microwave energy harvesting \cite{Fang2019} or frequency detection \cite{Jenkins2016,Louis2017}, and of reconstructing bio-inspired networks for neuromorphic computing \cite{Torrejon2017,Romera2018}. 
All these functionalities, often realized in the very same STNO devices, are based on basic spintronic phenomena, such as injection locking to an external rf signal \cite{Lebrun2015,Hamadeh2014}, synchronization of multiple STNOs \cite{Kaka2005,Locatelli2015,Lebrun2017,Tsunegi2018} or the spin torque diode effect \cite{Jenkins2016,Tulapurkar2005,Miwa2013,Naganuma2015}. 

Like any oscillator in nature, STNOs suffer from noise and their applicability in real practical devices must be measurable against their stability in terms of phase (or equivalently frequency). 
More specific to strongly nonlinear oscillators, such as STNOs, the amplitude noise 
must also be considered as it is converted into phase noise due to nonlinearity. In general, different mechanisms are found to govern the STNO's noise characteristics, usually expressed by its power spectral density (PSD). Principally, it includes thermal white noise processes, that dominate at higher offset frequencies from the carrier, and colored $1/f$ flicker noise processes, which dominate at low offset frequencies, i.e. at long timescales \cite{Wittrock2019_PRB}. So far, especially thermal noise effects have been under consideration and were studied both experimentally\cite{Quinsat2010,Grimaldi2017} and theoretically \cite{Kim2008,Slavin2009,Silva2010}. It was found that the oscillator's nonlinearity strongly affects the noise PSD and the power emission frequency spectrum in the thermal regime. However, the understanding of colored $1/f$ flicker noise and its impact on the oscillation behaviour is still limited and usually relies on a phenomenological treatment\cite{Wittrock2019_PRB} because of the noise's universality\cite{Rubiola2008} and its potential manifold origins. Sources of $1/f$ noise might be intrinsic such as fluctuations of magnetization (for magnetic sensor devices \cite{Diao2011,Nowak1999,Arakawa2012}), the incidence of defects and/or inhomogeneities in the magnetic layers or the tunnel barrier notably due to the fabrication process\cite{Eklund2014}. Note that in addition to intrinsic origins, external fluctuations of the driving dc current or the applied magnetic field might also play a role. In experiments, the existence of $1/f$ noise at low offset frequencies has been reported in the different types of spin torque nano oscillators based on the dynamics of a uniform magnetic mode \cite{Quinsat2010,Keller2010}, of the gyrotropic motion of a vortex core \cite{Grimaldi2017,Wittrock2019_PRB} or in nano-contact STNOs \cite{Eklund2014}. From the theoretical side, we have shown in a phenomenological model how the existence of both thermal noise and flicker noise is affecting the noise properties (phase and amplitude power spectral densities, PSDs) in the low offset frequency regime, and notably how the oscillator's nonlinearities play an important role \cite{Wittrock2019_PRB}.
%

In the present work, we investigate both theoretically and experimentally in spin transfer nano-oscillators (STNOs) how the presence of these different sources of noise, i.e. thermal and especially $1/f$ flicker noise, strongly impact the main oscillations' characteristics, in particular their spectral shape. More specifically, we study how the oscillation's spectral shape depends on the measurement duration and distinguish the correlation times of the different noise sources. We demonstrate how the spectral shape changes from a Lorentz shape at short measurement durations associated to white noise correlation, and to a Voigt -- or even Gaussian -- shape at longer durations with colored $1/f$ correlation. In complement to these experimental results, we develop a simulation scheme including both a basic flicker noise process and thermal fluctuations. Finally, we furthermore present a theoretical model (see section \ref{sec:theory}) in which the variance functions of the phase fluctuations are derived allowing to predict the shape of the frequency spectrum of STVOs.

\section{Methods}

\subsection{Experiments}

The experimental measurements presented in this work have been performed on vortex based spin transfer oscillators (STVOs). This type of STNOs exploit the gyrotropic motion of a magnetic vortex in a circular shaped nanodisk and convert its dynamics, which is sustained by spin transfer torque, into an electric rf signal due to the magnetoresistive effect\cite{Dussaux2010}. 

The studied samples consist of a pinned layer made of a conventional synthetic antiferromagnetic stack (SAF), a MgO tunnel barrier and a FeB free layer in a magnetic vortex configuration.
The  magnetoresistive ratio related to the tunnel magnetoresistance effect (TMR) lies around $100\,$\% at room temperature.  
The SAF is composed of \selectlanguage{ngerman} PtMn($15$)/""Co$_{71}$Fe$_{29}$($2.5$)/""Ru($0.86$)/""CoFeB($1.6$)/""Co$_{70}$Fe$_{30}$($0.8$) and the total layer stack is SAF/""MgO($1$)/""FeB($4$)/""MgO($1$)/""Ta/Ru, with the nanometer layer thickness in brackets.
\selectlanguage{english}
%
%
%
The circular tunnel junctions have an actual diameter of $2R = 375\,$nm. The presented experimental results have been recorded with an applied out-of-plane field of $\mu_0 H_{\perp} = 495\,$mT and a dc current of $I_{dc} = 5.5\,$mA. 
More details on the measurement techniques can be found in Ref. \cite{Wittrock2019_PRB}.

\subsection{Simulation}
\label{sec:methods-simulation}




We perform numerical simulations of the differential Thiele equation which describes the dynamics of the vortex core \cite{Thiele1973,Dussaux2012}:
\begin{gather*}
\boldsymbol{G} \times \frac{d\boldsymbol{X}}{dt} - \hat{D}(\boldsymbol{X}) \frac{d\boldsymbol{X}}{dt} 
- \frac{\partial W (\boldsymbol{X}, I_{\text{STO}}(t) )}{\partial \boldsymbol{X}} + \boldsymbol{F}^{STT} (\boldsymbol{X}, I_{\text{STO}}(t)  ) = 0  ~~~.
\end{gather*}
$\boldsymbol{X}$ denotes the vortex core position,  $\boldsymbol{G}$ the gyrovector, $\hat{D}$ the damping, $W$ the potential energy of the vortex, $\boldsymbol{F}^{STT}$ the spin-transfer force, and $I_{\text{STO}}$ the applied dc current. 

In this work, our objective is to include in the simulations both thermal and $1/f$ flicker noise. To do so, the contribution of the thermal noise is introduced through a fluctuating field varying the vortex core position \cite{Khalsa2015}. It follows a normal distribution with a zero mean value and a fluctuation amplitude given by $\Gamma$:
\begin{equation}
 \Gamma =  \dfrac{2k_B T D_0}{R^2 G}   ~~~,
 \label{ThieleEquation}
\end{equation}
\hspace{0.25cm} where  $k_B$ is the Boltzmann constant, $T$ corresponds to the temperature (set to 300 K in our simulations), $D_0$ is the linear (in the gyration amplitude) term of the damping force $\hat{D}$, $R$ is the nano-dot radius and $G$ is the amplitude of the gyroforce $\boldsymbol{G}$. The values of the parameters used in the explicit expression of the Thiele equation to simulate the vortex dynamics are given in the Supplementary Material \cite{supplementary_spectrum-with-flicker_2019}.

\begin{figure}[hbt!]
		\centering
		\vspace{1em}
		\includegraphics[width=0.54\columnwidth]{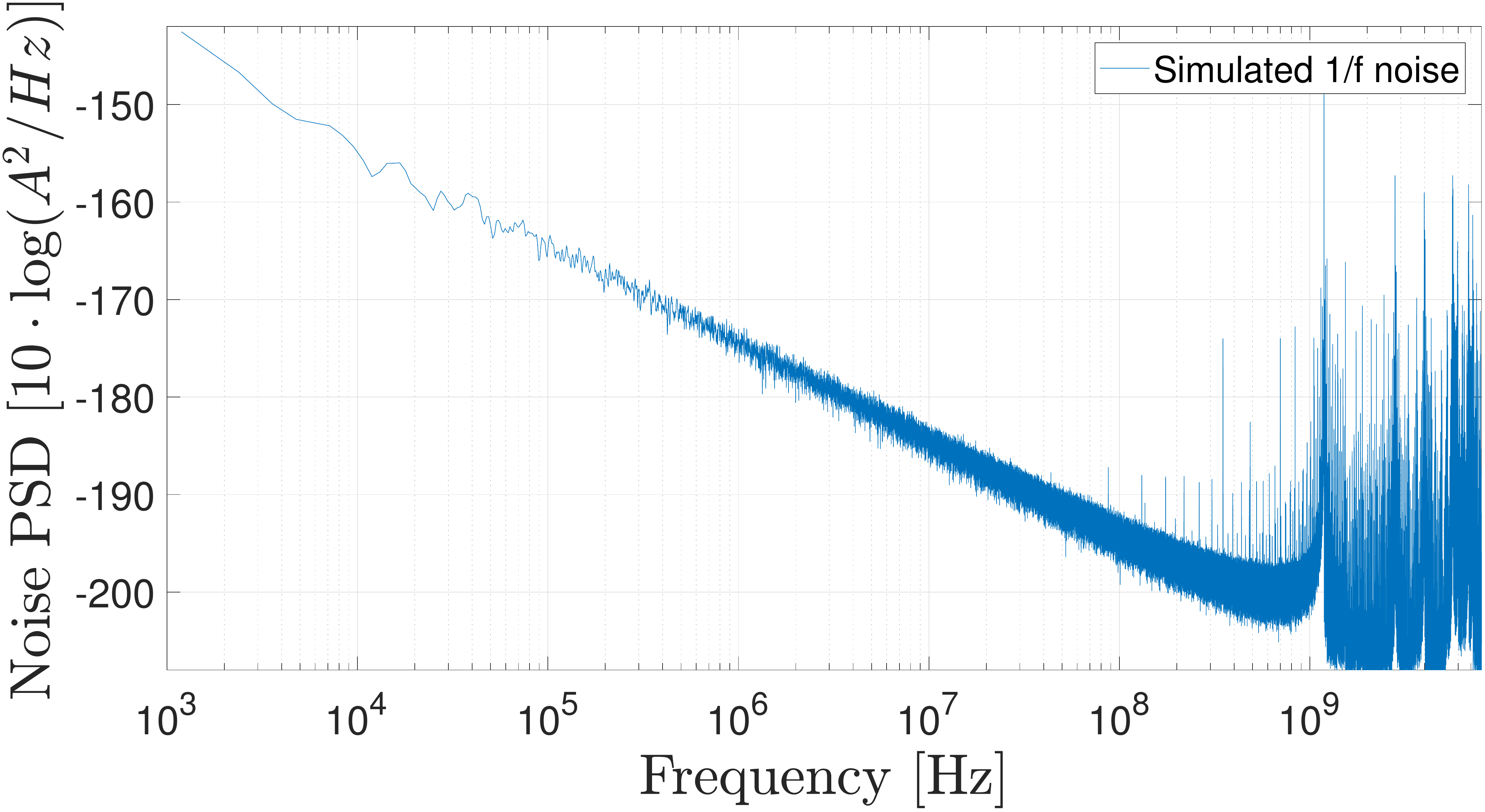}
		\caption[]{Modelled $1/f$ flicker noise PSD for the simulation, $\lambda=2.5\cdot 10^{-4}$. Fluctuations are added to the simulated dc current.}
		\label{fig_sim:generating-noise_simulation}
\end{figure}

As for the simulation of the flicker noise contribution, we introduce a random variable $r_{1/f}(t)$ that has a $1/f^1$ flicker noise PSD as presented in 
fig. \ref{fig_sim:generating-noise_simulation} (see Supplementary Material \cite{supplementary_spectrum-with-flicker_2019} for a more detailed description how this variable is constructed). This random process variable is then added to the applied dc current $I_{dc}$:
\begin{equation}
I_{STO}(t)=  I_{dc} + \lambda r_{1/f}(t)   ~~~.
\label{eq:simulation_ISTO_flicker}
\end{equation}

where $\lambda$ is a scalar factor chosen here to be $2.5\cdot 10^{-4}$ in order for the simulated amplitude and phase noise PSD to be close to measured ones. 

 

\section{Experimental noise PSD\lowercase{s} and frequency spectrum}
\label{sec:freq_spectrum}



\begin{figure}[tbh!]
		\centering
		\vspace{0em}
		\includegraphics[width=1\columnwidth]{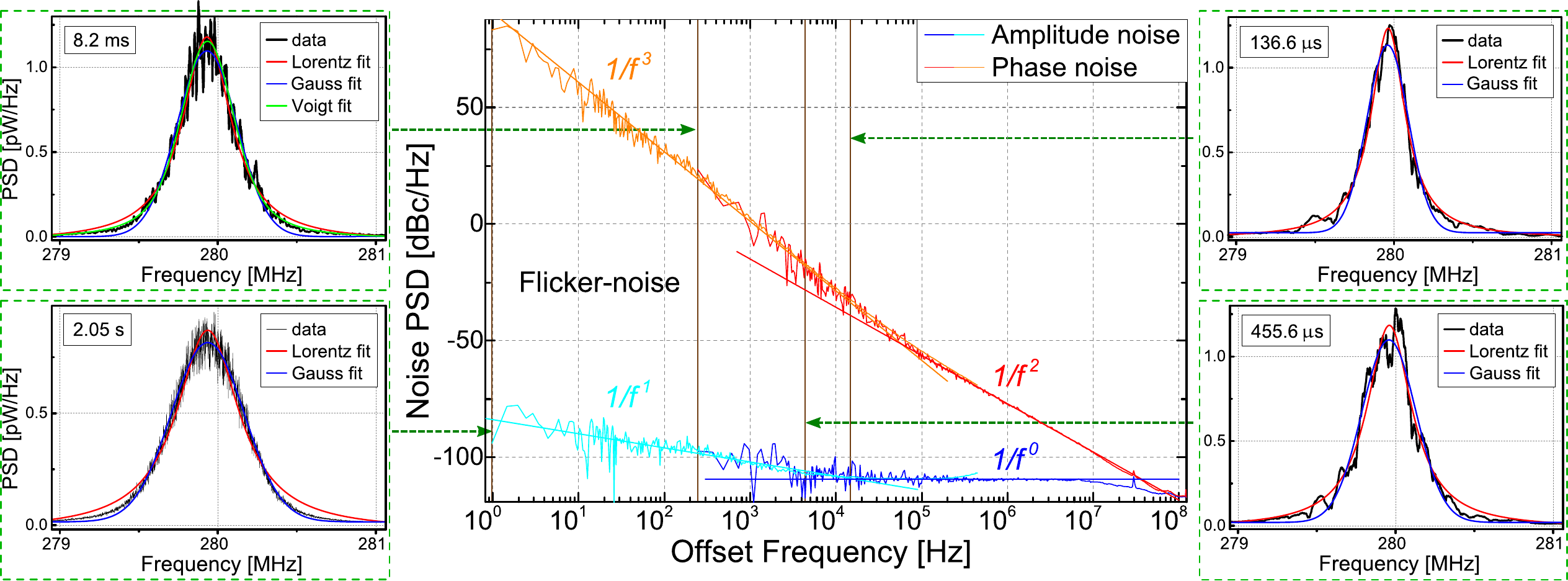}  
		\caption[]{Measurement of amplitude and phase noise ${S_{\delta \epsilon}}$ and ${S_{\delta \phi} }$, respectively, and of frequency spectra corresponding to different measurement durations.  $\mu_0 H_{\perp}=495\,$mT, $I = 5.5\,$mA.  The performed fits on the noise PSD show the characteristic $1/f^{\beta}$-behaviour for high (red, blue resp.) and low offset frequencies (orange, light blue resp.). } 
		\label{fig_exp:noise_and_spectra}
		\end{figure}

In the central panel of Fig. \ref{fig_exp:noise_and_spectra}, we present the measured noise PSD corresponding to amplitude (blue and cyan curves) and phase (red and orange curves) fluctuations. 
The different noise contributions can be directly identified due to their inverse power law behavior $PSD \sim 1/f^{\beta}$. At large offset frequencies  $f\gtrsim 10^5\,$Hz , the thermal noise is dominant and  the noise signature behaves linearly with exponent $\beta = 2$ for the phase (red curve in Fig. \ref{fig_exp:noise_and_spectra}) and $\beta=0$ for the amplitude noise (blue curve in Fig. \ref{fig_exp:noise_and_spectra}). Note that this description does not apply for frequencies above the relaxation rate frequency $f_p$ ($\approx 11.8\,$MHz here) above which the oscillator's nonlinearities strongly affect the noise PSDs  \cite{Kim2008,Grimaldi2017,Wittrock2019_PRB}.

At low offset frequency $f\lesssim 10^5\,$Hz, the noise behaviors are clearly different because the flicker noise becomes the dominant source of noise.  As seen in fig. \ref{fig_exp:noise_and_spectra}, the phase noise PSD can be fitted with $\alpha/f^3$ (see orange curve in  Fig\ref{fig_exp:noise_and_spectra}) and the amplitude noise with $\alpha/f^1$ (see cyan curve in Fig\ref{fig_exp:noise_and_spectra}).
In addition to the measured PSDs, we also present in Fig. \ref{fig_exp:noise_and_spectra} four frequency spectra recorded for different measurement times, i.e.  $T_{mes} \accentset{\wedge}{=} 2/f = 136.6\,\upmu$s,  $456.6\,\upmu$s,  $8.2\,$ms and  $2.05\,$s along with different fits on the measured spectra. 
We find that for short measurement times (typically $<1\,$ms), the power emission spectra can be fitted with an excellent agreement by Lorentzian curves. On the contrary, for longer measurement periods, the fits using Gaussian curves become more accurate with an excellent agreement for $2.05\,$s. In summary, we find a Lorentzian spectral shape in the regime of dominant thermal noise with the phase noise PSD $S_{\delta \phi}\sim 1/f^2$, and a Gaussian dominated shape in the regime of dominant flicker phase noise with $S_{\delta\phi}\sim 1/f^3$. Note that in the intermediate measurement time, the spectrum is described by a convolution of Lorentzian and Gaussian shapes, which is a Voigt curve shape, as exemplified for the  $8.2\,$ms measurement time in fig. \ref{fig_exp:noise_and_spectra}.



\floatsetup[figure]{style=plain,subcapbesideposition=top}
\begin{figure}[tbh!]
  \centering  
  \subfloat[]{  
  \includegraphics[width=0.46\columnwidth]{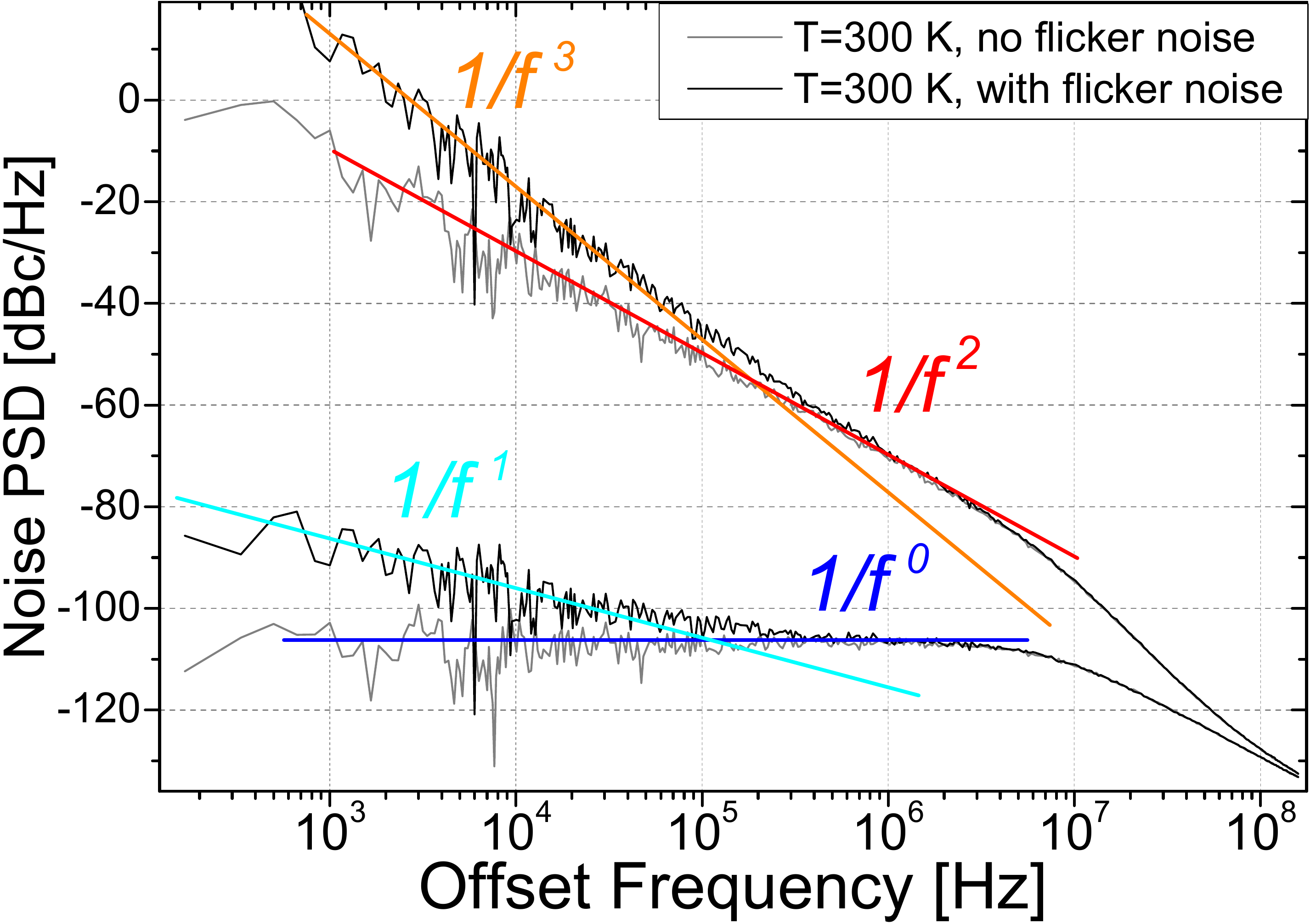} 
  \label{fig_sim:PhaseNoiseAmplitudeNoise} 
}  
  \subfloat[] 
  { 
 \hfill\hspace{-0.0cm} \includegraphics[width=0.465\columnwidth]{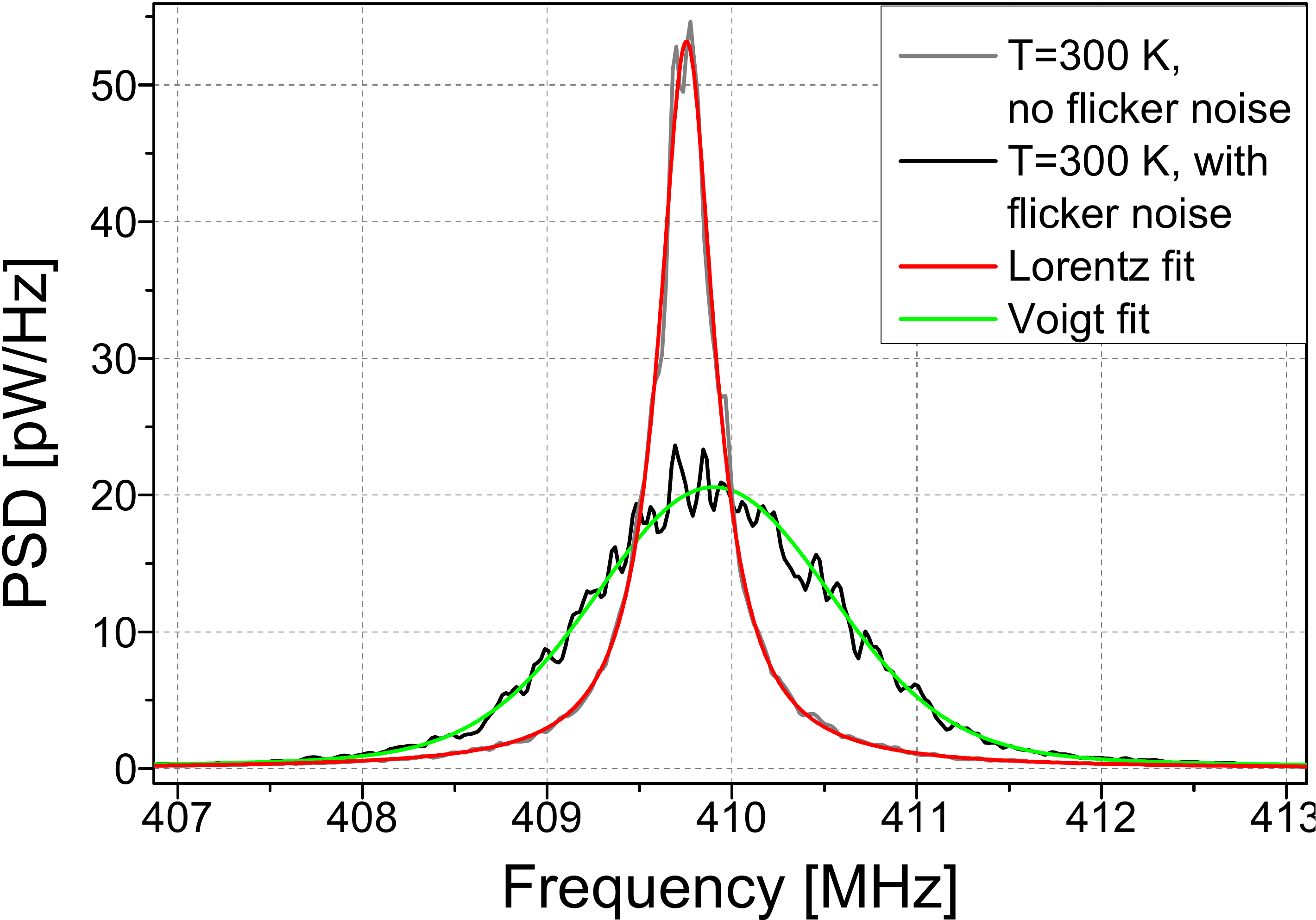}  
 \label{fig:spectrum_with-without_flicker_simulation}  
  }  
  \vspace{-0.0cm}
  \caption[]{Simulation results with and without flicker noise. (a) Amplitude and phase noise PSD. (b) Corresponding frequency spectra. }
  \label{fig_sim:noise_and_spectra}
\end{figure}

%
%

In fig. \ref{fig_sim:noise_and_spectra}, we display the noise PSD (top graph) and the frequency spectrum (bottom graph) obtained from the simulation of the vortex core dynamics (sec. \ref{sec:methods-simulation}) using an applied dc current of $4.5\,$mA far enough from the critical current for auto-oscillations $\sim 2\,$mA (thus, for the linewidth $2\cdot\Delta f_0 \ll f_p$). 
To investigate the impact of $1/f$ flicker noise on the spectral shape of the STVOs, a sufficiently large simulation time scale of $T_{\text{sim}}=10\,$ms is required. In fig. \ref{fig_sim:PhaseNoiseAmplitudeNoise}, we present two series of simulations, one with $\lambda = 2.5\cdot 10^{-4}$ (i.e. including thermal and flicker noise) and another one with $\lambda=0$ (i.e. including only thermal noise) (see eq. (\ref{eq:simulation_ISTO_flicker})).
At large offset frequencies, $\gtrsim 0.2\,$MHz, the simulated curves for the two $\lambda$ are equivalent as expected because the flicker noise contribution is negligible in this regime. Below this offset frequency, the two curves separate. For $\lambda = 0$ and offsets $<f_p \simeq 8\,$MHz, the noise PSD exhibits a plateau for the amplitude one and a $1/f^2$ decrease for the phase one. Note that in this case, despite the increase of phase noise  due to amplitude-to-phase noise conversion associated to STNO nonlinearities \cite{Kim2008,Slavin2009,Quinsat2010,Grimaldi2017,Wittrock2019_PRB}, the noise PSD characteristics considering only thermal noise can be treated as that of a linear oscillator. For $\lambda = 2.5\cdot 10^{-4}$ and offsets $<f_p \simeq 8\,$MHz, the simulated PSDs show the same trends and characteristics as the experimental ones presented in fig. \ref{fig_exp:noise_and_spectra}, thus validating our approach to describe the flicker noise in the vortex dynamics simulations by assuming a $1/f^1$ generating noise process in the dynamical equations.


In fig. \ref{fig:spectrum_with-without_flicker_simulation}, we depict the corresponding frequency spectrum of the oscillation power. For $\lambda = 0$, the fit using a Lorentzian curve (red curve) shows a perfect agreement with the simulation. On the contrary, when the flicker noise is taken into account (using $\lambda = 2.5\cdot 10^{-4}$), the spectrum becomes broader and its shape changes. The best fit is now obtained using a Voigt function (green curve), that is in fact close to be a Gaussian curve. We can thus conclude that the different noise sources contribute differently to the spectral shape of the oscillations. A direct consequence is that the spectral shape shall depend on the actual duration of the measurement as the flicker noise contribution becomes only significant at sufficiently large correlation times.

\begin{figure}[tbh!]
  \centering  
  \sidesubfloat[]
  {  
  \hspace{-0.0cm} 
  \includegraphics[width=0.426\columnwidth]{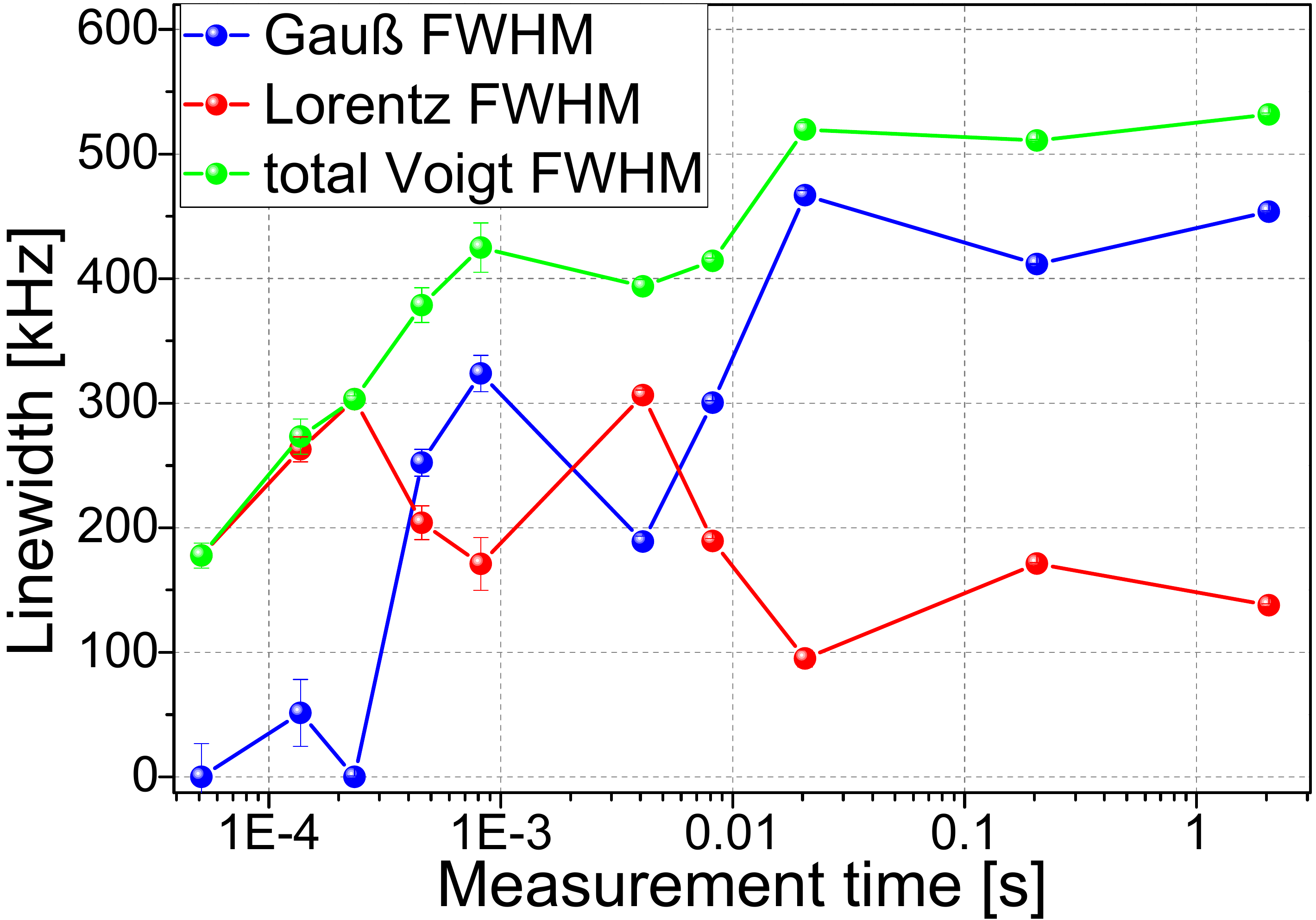} 
  \label{fig:Linewidth_lorentz-gauss-voigt_experiment} 
     }  
  \sidesubfloat[] 
  { 
 \hspace{-0.0cm} \includegraphics[width=0.426\columnwidth]{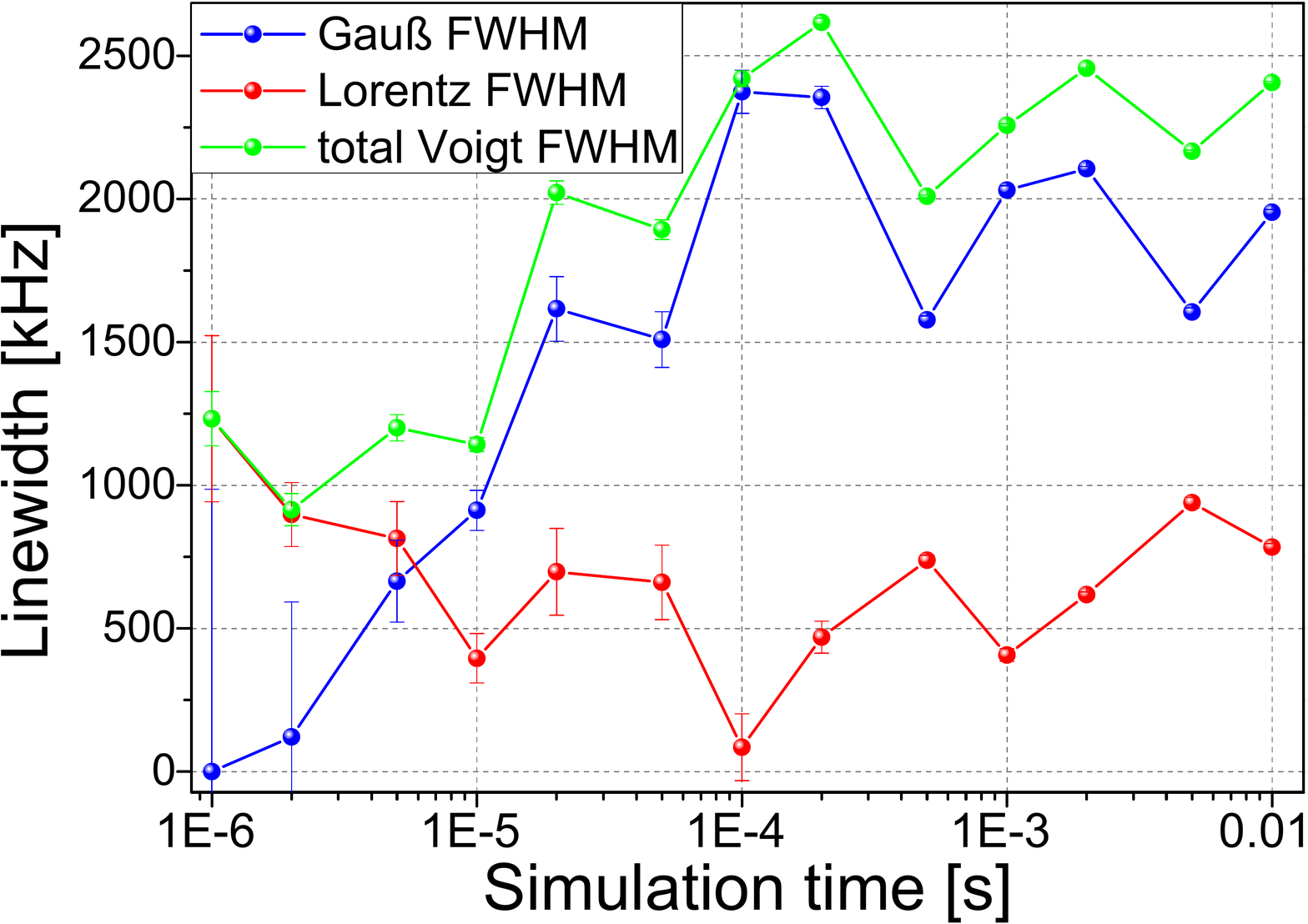} 
 \label{fig:Linewidth_lorentz-gauss-voigt_simulation}  
  }  
  \vspace{-0.0cm}
  \caption[]{Resulting fit parameters from the Voigt fits for (a) the measurement and (b) the simulation. Next to the total Voigt linewidth, the linewidth corresponding to convoluted Lorentz and Gauß shape is shown. }
  \label{fig:linewidth_lorentz-gauss-voigt}
\end{figure}

In fig. \ref{fig:linewidth_lorentz-gauss-voigt}, we plot the spectral linewidths of the according contributions (Lorentz, Gauss, Voigt) provided by the Voigt fits  as a function of the duration both for the experimental spectra (top graph) and the simulated ones (bottom graph). 
From these curves, we observe that the Lorentz contribution to the spectral linewidth does not change significantly with the recording time, whereas the total linewidth represented by the Voigt FWHM increases for higher measurement time, as well as the contribution from the Gaussian shape. 








\section{Theoretical model}  
\label{sec:theory}

Taking into account both thermal and flicker noise processes within the framework of the general nonlinear auto-oscillator theory \cite{Slavin2009},  the noise PSD of amplitude ${S_{\delta \epsilon}}$ and phase noise ${S_{\delta \phi}}$ can be written using the following expressions we recently described in  \cite{Wittrock2019_PRB}:

\begin{align}
{S_{\delta \epsilon}} &= \frac{\Delta f_0}{\pi} \cdot \frac{1}{f^2 + f_p^2} + \frac{1}{4 p_0 \pi^2 \left( f_p^2 + f^2   \right)} \cdot  \frac{\alpha_{\delta \epsilon}}{f^{\gamma}} \label{eq:all-noise_amp} \\
{S_{\delta \phi} } &=  \frac{\Delta f_0}{\pi f^2} + \frac{1}{4 p_0 \pi^2 f^2 } \cdot \frac{\alpha_{\delta \phi}}{f^{\gamma}} + \frac{\nu^2 f_p^2}{f^2} {S_{\delta \epsilon}}  ~~~,   \label{eq:all-noise_phase}
\end{align}

where $\Delta f_0$ is the linear oscillation linewidth and $f_p$ is the characteristic frequency of relaxation back to the stable oscillation power $p_0$ after small perturbations $\delta p$. The parameter $\nu = N p_0/(\pi f_p)$ with $N=d\omega(p)/dp$ the nonlinear frequency shift coefficient is the normalized dimensionless nonlinear frequency shift and quantifies the coupling between phase and amplitude due to nonlinearity.  The parameter $\alpha_{\delta x}$ describes the generating flicker noise process acting differently on amplitude/phase $x$. Its characteristic exponent $\gamma$ has been found $\gamma = 1$. 

In the following, we focus on the theoretical description of the phase noise ${S_{\delta \phi} }$, that is the most important for describing the spectral shape of the oscillation, which is one of the main objectives of this study.
In eq. (\ref{eq:all-noise_phase}), the three terms describe different contributions to the phase noise. The linear contributions are the ones being independent of $\nu$, whereas the nonlinear ones are scaling with $\nu^2$, converting amplitude ${S_{\delta \epsilon}}$ into phase noise ${S_{\delta \phi} }$. Moreover, all terms proportional to $\Delta f_0$ describe the thermal noise, and the terms proportional to $\alpha_{\delta x}$ the flicker noise contribution. 
For simplicity, we choose ${S_{\delta \phi} }$ to be expressed in terms of the angular frequency $\omega = 2\pi f$ and introduce the simplified parameters $\alpha_{\text{ph}}$ and $\alpha_{\text{amp}}$. 
Thus in the following, we evaluate the phase noise in its simplified form: 
\begin{align}
{S_{\delta \phi} } = \frac{2\Delta \omega_0}{\omega ^2}  +  \nu^2 \frac{2\Delta\omega_0}{\omega^2} \frac{\omega_p^2}{(\omega^2+\omega_p^2)} \nonumber \\ 
+ \frac{\alpha_{ph}}{\omega^3} + \frac{ \alpha_{amp}}{ (\omega_p^2+\omega^2) \omega^3}  ~~~.   \label{eq:PN_simplified}
\end{align}
The parameters $\alpha_{\text{ph}}$ and $\alpha_{\text{amp}}$ can be identified with the experimental magnitudes:
\begin{align*}
 \alpha_{\text{ph}} &= (2\pi )^3 \alpha_{\text{ph,exp}}= 2\pi    \alpha_{\delta \phi}/p_0 \\
 \alpha_{\text{amp}} &= (2\pi )^3 \omega_p^2 \alpha_{\text{amp,exp}} = 2\pi  \omega_p^2 \nu^2  \alpha_{\delta \epsilon}/p_0  ~~~,
 \end{align*}
\hspace{0.25cm} where $\alpha_{\text{amp,exp}}$ and $\alpha_{\text{ph,exp}}$ are the fitting parameters $\alpha_{x\text{,exp}}/f^3$ on the nonlinear low offset frequency flicker phase noise converted from amplitude noise (last term in eq. (\ref{eq:all-noise_phase})) and on the linear pure phase flicker noise (2nd term in (\ref{eq:all-noise_phase})) respectively\cite{Wittrock2019_PRB}.

From eq. (\ref{eq:PN_simplified}), we can compute the variance of the oscillator's phase fluctuations \cite{supplementary_spectrum-with-flicker_2019,Demir2002,Demir2006,Godone2008,Makdissi2010} $\Delta\phi^2$:  
\begin{align}
\frac{\Delta\phi^2}{2}  &=   \frac{1}{\pi} \int\limits_{0}^{\infty} \left( 1 - \cos( \omega t) \right) \cdot S_{\delta \phi }  d\omega      ~~~. \label{eq:noise-PSD_2}
\end{align}

Assuming $S_{\delta\epsilon} \ll S_{\delta\phi}$ and a stationary ergodic process, which is gaussian distributed (via the central limit theorem), the signal's autocorrelation can be approximated by \cite{supplementary_spectrum-with-flicker_2019}:
\begin{align}
K(t) \approx p_0 e^{-i\omega t} e^{-\Delta\phi^2/2}  ~~~.   \label{eq:autocorrelation_def}
\end{align}

Then applying the Wiener-Khintchine theorem, the frequency spectrum's PSD of the signal $x$ can be calculated through the Fourier transform of the signal's autocorrelation:
\begin{align}
S_x (\omega) = \int\limits_{\mathbb{R}} K_x(t) e^{i\omega t} dt  ~~~.  \label{eq:wiener-khintchine}
\end{align}

\subsection{Variance of phase fluctuations}
\label{subsec:variance-calculation}


Following the expression of eq. (\ref{eq:noise-PSD_2}) for the computation of the fluctuation's variance calculation, we first give the expression for the variance of the thermal noise contributions (all terms proportional to $\Delta \omega_0$ in eq. (\ref{eq:PN_simplified})):


\begin{align*}
 \begin{array}{>{\displaystyle}c>{\displaystyle}l>{\displaystyle}l} \left( \frac{\Delta\phi^2}{2} \right)_{th} &\approx  \frac{1}{\pi}\cdot \displaystyle\int\limits_{\omega_c}^{\infty} \left( 1-\cos(\omega t) \right) S_{\delta \phi , th} d\omega  &  \\[3ex]
  &=      \frac{\alpha_l}{\pi\omega_c} + \frac{\alpha_l}{\pi} \cdot \left( \frac{ t \pi}{2}  -   t\cdot  \Si(t\omega_c) - \frac{ \cos(\omega_c t)}{\omega_c}    \right) & \multirow{1}{*}[0.3em]{  $\left.\vphantom{\int\limits_2^2}\right\}$\text{linear part} }  \\[3ex]
&+ \frac{\alpha_{nl}}{\pi}  \left( \frac{\omega_p + \omega_c \arctan(\frac{\omega_c}{\omega_p})}{\omega_c \omega_p^3}   -   \frac{\pi}{2\omega_p^3}      \right) &  \multirow{2}{*}[0.5em]{  $\left.\vphantom{\begin{matrix} a\\a\\a\\a\\a\\a\\a\\a\\\end{matrix}}\right\}$\text{nonlinear part} }   \\[3ex]
&- \frac{\alpha_{nl}}{2\pi\omega_p^3} \left[ \sinh(\omega_p t) \cdot \pi - \omega_p t \pi - \sinh(\omega_p t) \left( \Si\left(t(\omega_c + i\omega_p)\right) + \Si\left(t(\omega_c - i\omega_p)\right) \vphantom{\frac{x^2}{x^2}}    \right)     \vphantom{\frac{x_c^2}{x_c^2}}   \right.   &  \\[3ex]
&\left. + i \cosh(\omega_p t) \left( \Ci\left(t(\omega_c + i\omega_p)\right) - \Ci\left(t(\omega_c - i\omega_p)\right) \vphantom{\frac{x^2}{x^2}}    \right)                      \vphantom{\frac{x_c^2}{x_c^2}}   \right.    \left.-  2\omega_p t \Si(t\omega_c)  +  \frac{2\omega_p \cos(t\omega_c)}{\omega_c}                \vphantom{\frac{x_c^2}{x_c^2}}   \right]   &   \\[3ex]
\end{array} \\~~~,
\end{align*}
\hspace{0.25cm} where we insert for the linear thermal noise part $\alpha_l \coloneqq 2\Delta\omega_0$ and $\alpha_{nl} \coloneqq 2\Delta\omega_0 \nu^2 \omega_p^2$. $\Si(x)$ and $\Ci(x)$ denote the sine and cosine integral respectively, and $\omega_c$ is a lower frequency cutoff.
 For the two flicker noise terms, we calculate:
\begin{align*}
 \begin{array}{>{\displaystyle}c>{\displaystyle}l>{\displaystyle}l} \left( \frac{\Delta\phi^2}{2} \right)_{1/f} &=      \frac{ \alpha_{ph}}{\pi} \left[ \frac{1}{2\omega_c^2} + \frac{t^2}{2} \left(   \frac{\sin(\omega_c t)}{t\omega_c} - \Ci(t\omega_c) - \frac{\cos(\omega_c t}{(\omega_c t)^2} \right) \right] & \left.\vphantom{\frac{t^2\pi}{4}}\right\}\text{pure flicker} \\[3ex]
 &+ \frac{\alpha_{amp}}{\pi} \left[ - \frac{\ln(\omega_p^2+\omega_c^2)}{2\omega_p^4}  + \frac{1}{2\omega_c^2\omega_p^2} + \frac{\ln(\omega_c)}{\omega_p^4}     \right] &    \multirow{4}{*}[0.5em]{  $\left.\vphantom{\begin{matrix} a\\a\\a\\a\\a\\a\\a\\a\\a\\a^2\\ \end{matrix}}\right\}$\text{nonlinear flicker} }    \\[3ex]
 &+ \frac{\alpha_{amp}}{2 \pi \omega_p^4} \left[  i \sinh(\omega_p t)\cdot \left[ \Si\left(\omega_c t + i \omega_p t\right) - \Si\left(\omega_c t - i \omega_p t\right)  \right]          \vphantom{\frac{\omega_p^2 \sin(\omega_c t)}{\omega_c^2}} \right.  &  \\[3ex]
 & \left. \hphantom{+ \frac{\alpha_{amp}}{2 \pi \omega_p^4} } + \cosh(\omega_p t) \cdot \left[  \Ci\left(\omega_c t + i \omega_p t\right)  +     \Ci\left(\omega_c t - i \omega_p t\right) \right]  \right. & \\[3ex]
 &  \left.  \hphantom{+ \frac{\alpha_{amp}}{2 \pi \omega_p^4} }  - \Ci(\omega_c t) \omega_p^2 t^2 - 2\Ci(\omega_c t) + \frac{\omega_p^2 t \sin(\omega_c t)}{\omega_c}  -  \frac{\omega_p^2\cos(\omega_c t)}{\omega_c^2}       \vphantom{\begin{matrix} a \\ a \\  \end{matrix}}    \right]   &
 \end{array} \\~~~.
\end{align*} 

In all the calculation above, we have assumed a practical lower frequency cutoff $\omega_c$, which is necessary to circumvent the integrals' divergence at $\omega = 0$. 
Note that for a comparative analysis of the timing jitter and frequency stability, the divergence is also often avoided by adding a filter function to eq. (\ref{eq:noise-PSD_2}) \cite{Makdissi2010}, invoking different variance definitions, as e.g. the Allan, modified Allan, or Hadamard variance \cite{Riley2008-handbook_NIST}.
However, it is to be emphasized that in real physical systems, our assumption will remain valid because the measurement time is usually restricted and the bandwidth is finite. In this sense, $\omega_c$ reflects the measurement duration studied in section \ref{sec:freq_spectrum}. 
Moreover, the autocorrelation's delay $t$ has to be limited, too, as $\omega_c t < 2\pi$.
A reasonable approximation for the variance function is therefore $\omega_c t \ll 1$. 
Furthermore assuming $\omega_c \ll \omega_p$, we can approximate the above expressions.
For the thermal noise part, with the analytical expressions for the prefactors  $\alpha_l = 2\Delta\omega_0$ and $\alpha_{nl} = 2\Delta\omega_0 \nu^2 \omega_p^2$ introduced above, this gives:
\begin{align}
\left( \frac{\Delta\phi^2}{2} \right)_{th} = \Delta\omega_0 \left[ (1+\nu^2) t - \nu^2 \frac{1-e^{-\omega_p t}}{\omega_p} \right]   ~~~.  \label{eq:thermal_variance_Tiberkevich}
\end{align}
This representation agrees with the one given by Tiberkevich \etal in Ref. \cite{Tiberkevich2008}, where also a detailed discussion of the temperature related effects in STNOs can be found. As the phase variance varies linearly only for time intervals $t\gg 2/\omega_p$, the STNO's power spectrum is thus in general non-Lorentzian.
However, if the generation linewidth $\Delta\omega$ is sufficiently small, i.e. $2\Delta \omega \ll \omega_p$, the exponential factor in (\ref{eq:thermal_variance_Tiberkevich}) can be neglected at a typical decoherence time $t \sim 1/\Delta\omega$:
\begin{align*}
\left( \frac{\Delta\phi^2}{2} \right)_{th} \approx \Delta\omega_0 (1+\nu^2)t - \frac{\nu^2}{2\omega_p} ~~~.
\end{align*}
This leads to a linear variance and thus, a Lorentzian power spectrum with a $\text{FWHM}$ of:
\begin{align*}
\text{FWHM} = 2\Delta\omega = 2\Delta\omega_0 (1+\nu^2) ~~~.
\end{align*}
If the power fluctuations' correlation time holds $\tau_p = 2/\omega_p \gg 1/\Delta\omega$ and is therewith much longer than the oscillator's coherence time, eq. (\ref{eq:thermal_variance_Tiberkevich}) yields:
\begin{align*}
\left( \frac{\Delta\phi^2}{2} \right)_{th} \approx \Delta\omega_0 \left(  t + \frac{\nu^2 \omega_p t^2 }{2}   \right)  ~~~,
\end{align*}
\hspace{0.25cm} which is quadratic and hence leads to a Gaussian power spectrum.
However, in the case of large amplitude steady state oscillations of the STVO as investigated in this work, the condition $2\Delta \omega \ll \omega_p$ is always fulfilled. 
For the flicker noise, we obtain in approximation:
\begin{align}
\left( \frac{\Delta\phi^2}{2} \right)_{1/f} 
&= \frac{ \alpha_{ph}}{\pi} \frac{t^2}{2} \left[     \frac{3}{2} - \ln(\omega_c t) - \gamma_{EM}   \right]  \nonumber  \\
&+ \frac{\alpha_{amp}}{2 \pi \omega_p^2} \left[  \frac{\ln\left( \nicefrac{\omega_c^2}{\omega_p^2} \right)}{2\omega_p^2} \right.   \nonumber  
 \left.  -  \left(t^2 + \nicefrac{2}{\omega_p^2} \right) \left( \ln(\omega_c t) + \gamma_{EM} \right)   \vphantom{\frac{\ln\left( \nicefrac{\omega_c^2}{\omega_p^2} \right)}{2\omega_p^4} }  +    2 t^2  \vphantom{\frac{\ln\left( \nicefrac{\omega_c^2}{\omega_p^2} \right)}{2\omega_p^4} }  \right]    ~~~,  \label{eq:variance_flicker_approx}
\end{align}
\hspace{0.25cm} where we assume that $t \gg 2/\omega_p$ 
 and introduce the Euler-Mascheroni constant $\gamma_{EM}\approx 0.58$.

\begin{figure}[bth!]
  \centering  
     \captionsetup[subfigure]{} 
	\raisebox{-0.26cm} {  
\sidesubfloat[]{       
  \includegraphics[width=0.4462\columnwidth ]{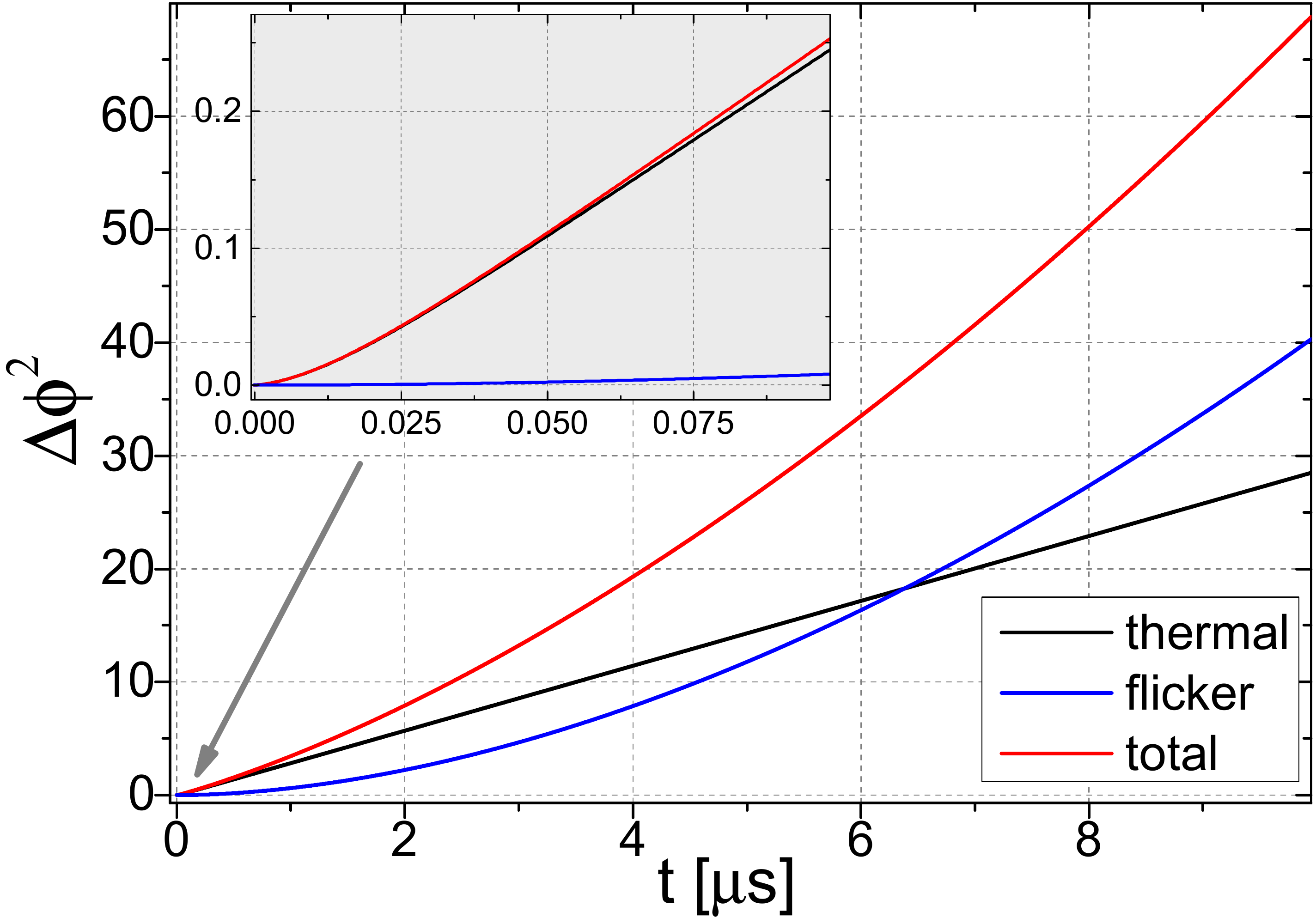}\label{fig_theo:variances}  
}  }
\captionsetup[subfigure]{} 
  \sidesubfloat[] 
  { 
 \hspace{-0.0cm} \includegraphics[width=0.4462\columnwidth]{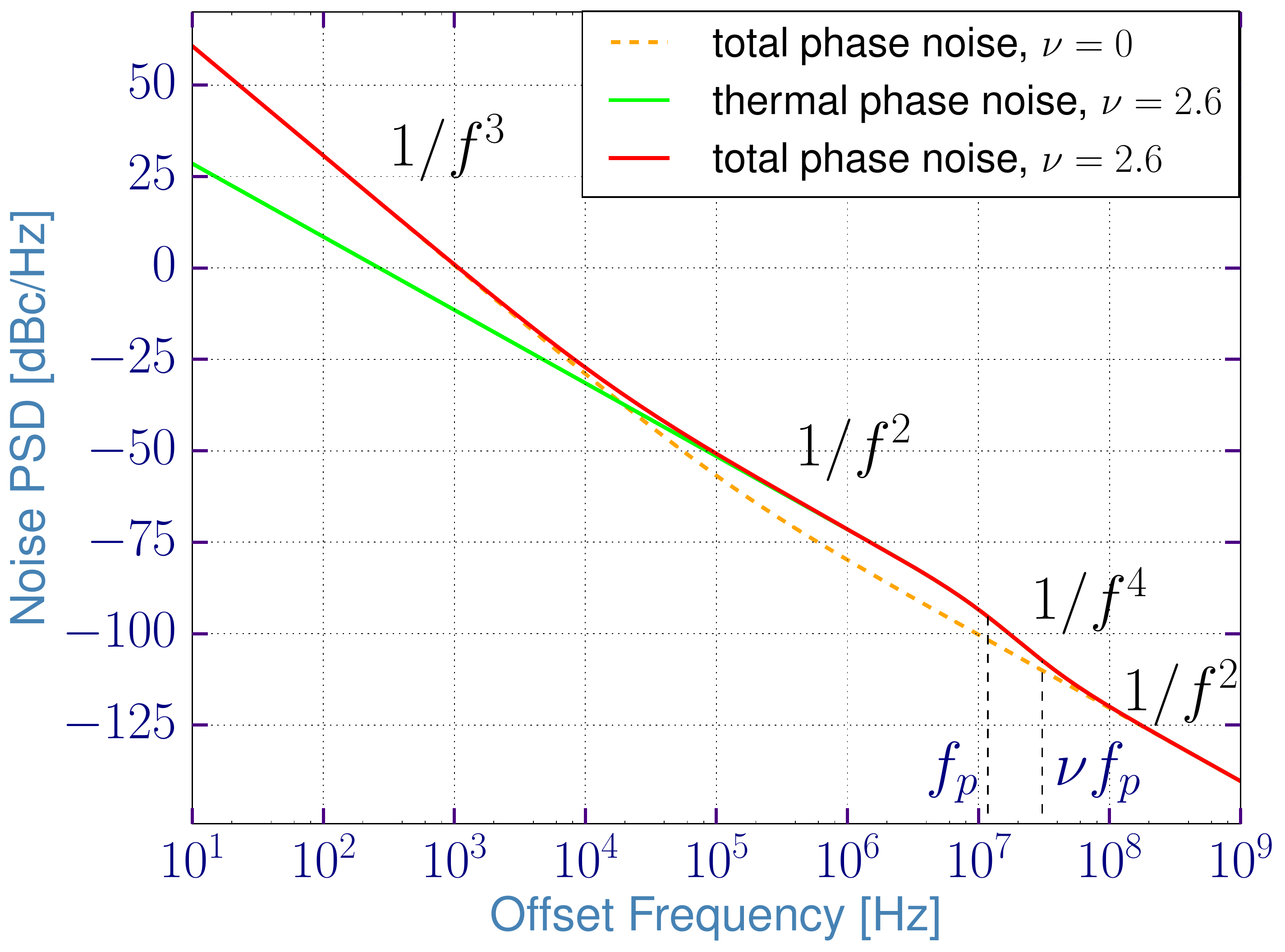}\label{fig_theo:noise} 
  }
  \caption[]{ \textbf{(a)} Variance functions vs. time $t$. The inset shows the same functions at small times. \textbf{(b)} Corresponding phase noise PSD as a function of frequency offset. The curves named ''total'' include both thermal and flicker noise contributions. }
  \label{fig_theo:variances_and_noise}
\end{figure}

In fig. \ref{fig_theo:variances}, we plot the variance functions for the thermal and the flicker noise contributions using the experimental parameters of the STVO studied in section \ref{sec:freq_spectrum} exhibiting a linewidth $\text{FWHM} = 2 \Delta f_0\cdot (1+\nu^2) = 450\,$kHz, a relaxation frequency $f_p=11.8\,$MHz, and a nonlinearity parameter $\nu=2.6$. Corresponding to a standard measurement time, the frequency cutoff is set to $f_c=500\,$Hz. The fitting parameters of the linear and nonlinear flicker noise in the sample are $\alpha_{\text{ph,exp}}=1.15\cdot 10^9$ and $\alpha_{\text{amp,exp}}=2.5\cdot 10^7$ respectively.

At small times $t$, displayed in the inset of figure \ref{fig_theo:variances}, we clearly see that the thermal noise contribution (black curve) is the dominant one to the variance function. As already mentioned, this is nonlinear for very small times before it becomes almost linear for increasing $t$. However, at even higher $t$ ($\gtrsim 7\,\upmu$s in graph \ref{fig_theo:variances}), because the flicker noise contribution starts to dominate, the total variance function becomes again nonlinear as the flicker variance appears to be almost quadratic. 

In fig. \ref{fig_theo:noise}, we plot the corresponding phase noise power spectral density based on equation (\ref{eq:all-noise_phase}). 
Similar to the noise PSD shown in section \ref{sec:freq_spectrum}, we again emphasize the dominance of thermal noise at higher offset frequencies and of $1/f$ flicker noise at the lower ones ($f\lesssim 10^5\,$Hz).
 

\subsection{Theoretical frequency power spectrum}

\begin{figure}[bth!]
  \centering  
  \captionsetup[subfigure]{}  
  \sidesubfloat[]{  
  \hspace{-0.14cm} 
  \includegraphics[width=0.382\columnwidth]{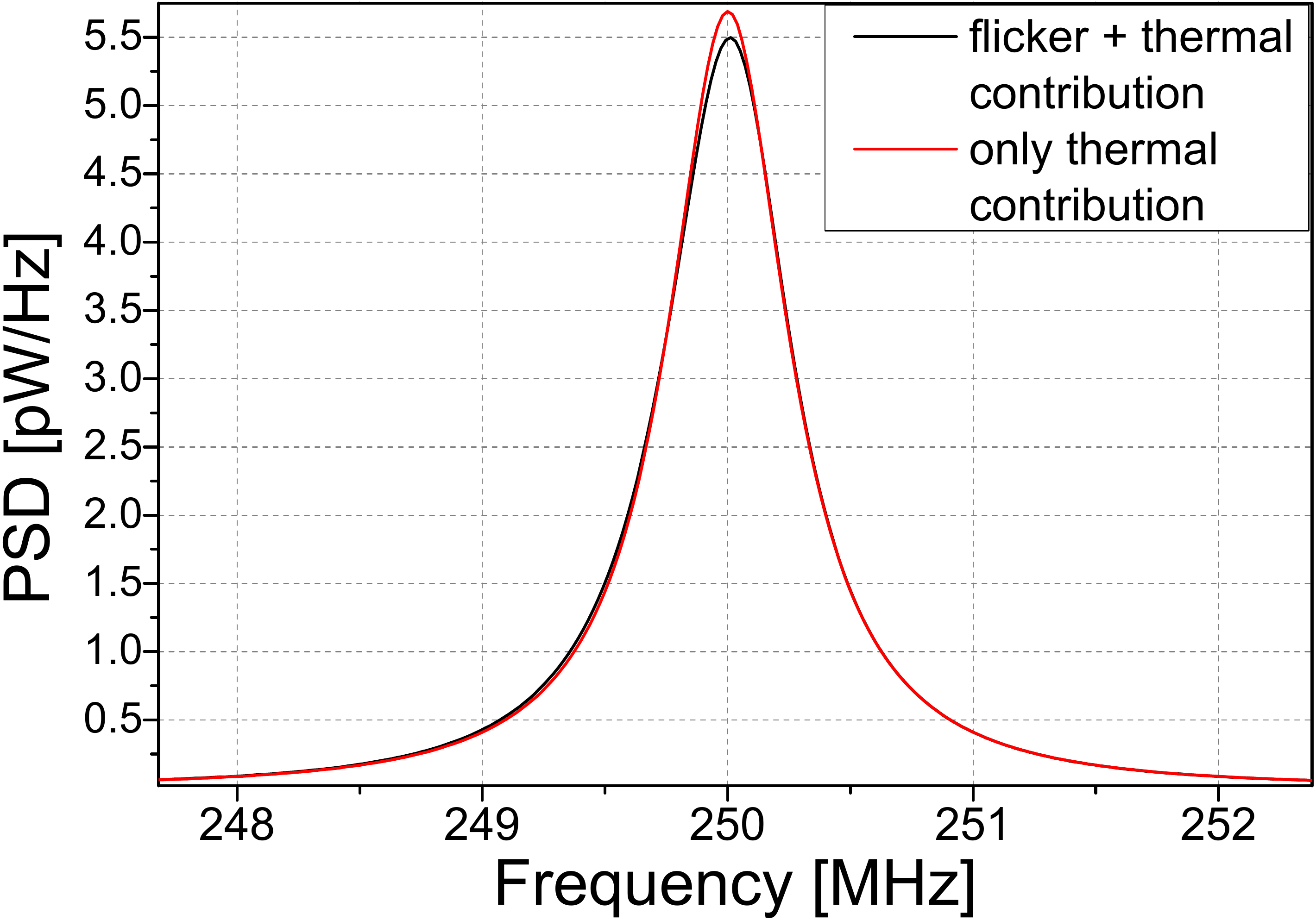}\label{fig_theo:spectrum-1} 
     } 
  \sidesubfloat[] 
  { 
 \hspace{-0.1cm} \includegraphics[width=0.382\columnwidth]{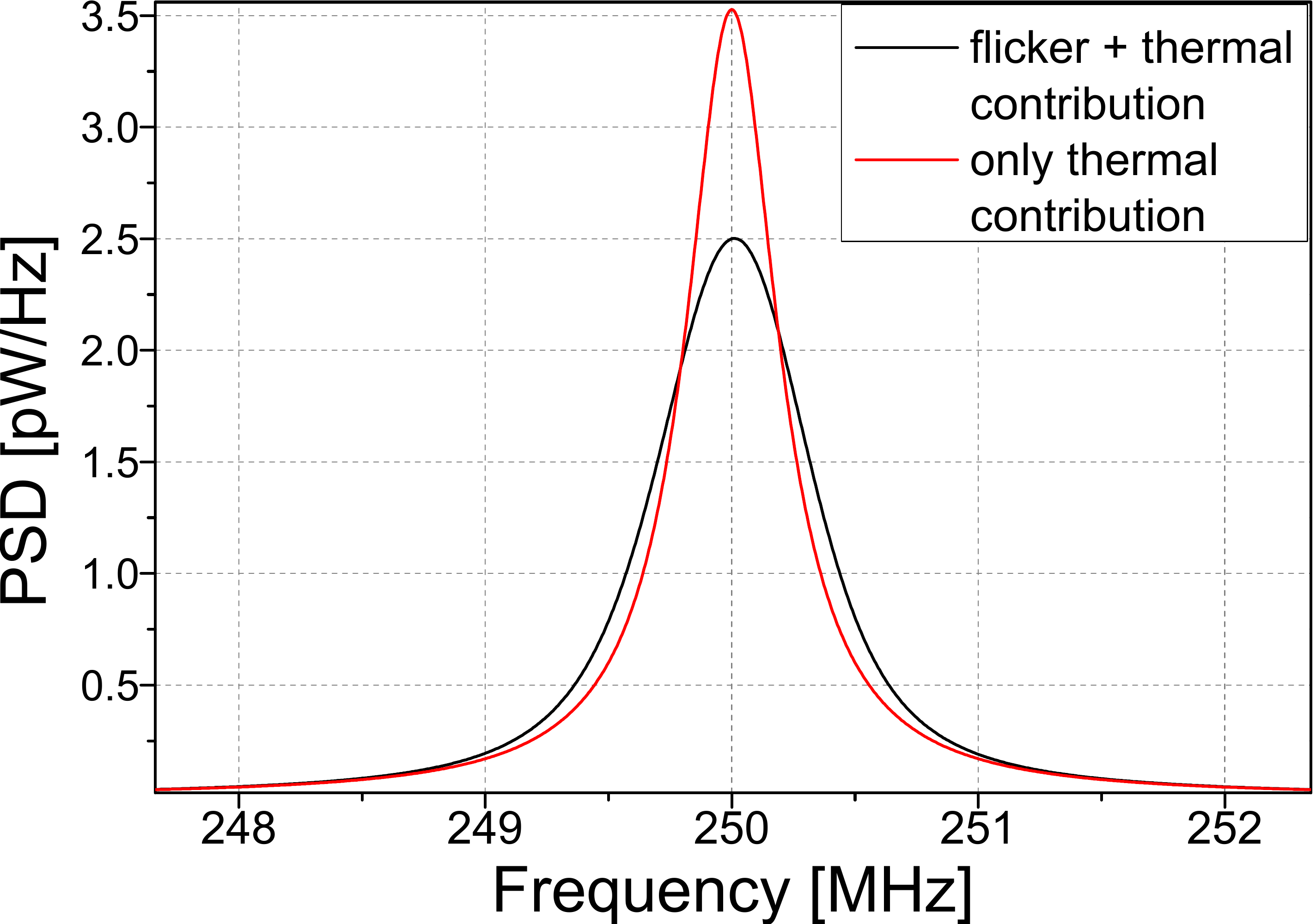}\label{fig_theo:spectrum-2}  
  }  
  \vspace{-0.0cm}
  \caption[]{ Calculated frequency spectra of the STNO signal with \textbf{(a)} $f_c = 10^5\,$Hz  and \textbf{(b)} $f_c = 1\,$Hz.  }
  \label{fig_theo:spectra}
\end{figure}

In fig. \ref{fig_theo:spectra}, we present the calculated power spectrum using the expressions derived from the theoretical model. From the calculated variance $\Delta \phi^2$ of the phase fluctuations, we can determine the power spectrum according to eqs. (\ref{eq:autocorrelation_def}) and (\ref{eq:wiener-khintchine}).

As done for the simulations, we can now compare the theoretical spectral shape in the two cases i.e. only the thermal noise is considered or both thermal and flicker contributions are taken into account. For these calculations, we have used the same STVO parameters as before and only the frequency-cutoff $f_c$ is changed. In fig. \ref{fig_theo:spectrum-1}, we observe that for $f_c=10^{5}\,$Hz (fig. \ref{fig_theo:spectrum-1}), the  spectrum is almost equivalent to the one only taking into account the thermal noise. In this case, the spectrum exhibits a Lorentzian shape due to its quasi-linear variance function $\Delta \phi^2$. On the contrary, for $f_c=1\,$Hz (fig. \ref{fig_theo:spectrum-2}), corresponding to a longer measurement time, the two spectra clearly differ. While the spectral shape associated to pure thermal noise is, as expected, still of Lorentzian type, the power spectrum in the presence of flicker noise is more complex. As its variance function shown in fig. \ref{fig_theo:variances} is nearly quadratic, we find a convolution of Lorentzian and Gaussian shapes, that is a Voigt function. This result obtained from the theoretical model thus reproduces excellently what we have previously shown both in the experiments and the simulations. 

\section{Conclusion}

In this work, we measure the noise characteristics of vortex based spin torque oscillators and observe that they are dominated by thermal noise at large offset frequencies and by flicker noise mechanisms at lower ones  $f\lesssim 10^5\,$Hz. In order to analyze these results, we perform simulations of the oscillator's noise properties by including the thermal contribution and as well, more originally, the flicker noise processes existing in the vortex dynamics in STVOs. To this aim, we have used the differential Thiele approach \cite{Thiele1973} together with a $1/f^1$ shaped generated noise. An important outcome of the simulations is that the presence of a simulated flicker noise clearly modifies the actual spectral shape of the output signal. Being almost purely Lorentzian with only thermal noise, the spectral shape becomes rather Gaussian in the presence of flicker noise. This behavior is indeed precisely the one we observed experimentally by recording power spectra using different measurement times. These results are then corroborated to a theoretical model that we have developed applying the nonlinear auto-oscillator theory including not only thermal but also flicker noise processes. Doing so, we succeed to derive the complete phase fluctuation's variance function and in consequence the theoretical shape of the frequency spectrum. In complete agreement with both experiments and simulations, we find that because of the different noise type correlation times, the STNO's spectral shape indeed dependends on the measurement duration, being Lorentzian type on short time scales and becoming Voigt type at longer ones. We believe that these findings are especially important regarding the anticipated diverse applications of STNOs, particularly if frequency stability is required on long time scales. Moreover, because the approach used to described the influence of thermal and flicker noise in presence of nonlinearities is not restricted to the description of STNOs \cite{Slavin2009}, the predictions made on the consequences on the spectral shape of the power spectra might also be valid for any other type of nonlinear oscillators that can be found in Nature.

%
%

\section{Acknowledgment}

S.W. acknowledges financial support from Labex FIRST-TF under contract number ANR-10-LABX-48-01. 
P.T. acknowledges   support under   the   Cooperative   Research   Agreement   Award No.  70NANB14H209,  through  the  University  of  Maryland. 
The work is supported by the French ANR project ''SPINNET'' ANR-18-CE24-0012.

\bibliography{literatur_promo}

\end{document}



\title{Influence of flicker noise and nonlinearity on the frequency spectrum of spin torque nano-oscillators  \\
 -- Supplementary -- }


\author{Steffen Wittrock}\email[]{steffen.wittrock@cnrs-thales.fr}\affiliation{Unit\'{e} Mixte de Physique CNRS/Thales, Univ. Paris-Sud, Univ. Paris-Saclay, 1 Avenue Augustin Fresnel, 91767 Palaiseau, France }
\author{Philippe Talatchian}\affiliation{Unit\'{e} Mixte de Physique CNRS/Thales, Univ. Paris-Sud, Univ. Paris-Saclay, 1 Avenue Augustin Fresnel, 91767 Palaiseau, France }\affiliation{Institute for Research in Electronics and Applied Physics, Univ. of Maryland, College Park, MD, USA}
\author{Sumito Tsunegi}\affiliation{National Institute of Advanced Industrial Science and Technology (AIST),
Spintronics Research Center, Tsukuba, Ibaraki 305-8568, Japan}
\author{Denis Crété}\affiliation{Unit\'{e} Mixte de Physique CNRS/Thales, Univ. Paris-Sud, Univ. Paris-Saclay, 1 Avenue Augustin Fresnel, 91767 Palaiseau, France }
\author{Kay Yakushiji}\affiliation{National Institute of Advanced Industrial Science and Technology (AIST),
Spintronics Research Center, Tsukuba, Ibaraki 305-8568, Japan}
 \author{Paolo Bortolotti}\affiliation{Unit\'{e} Mixte de Physique CNRS/Thales, Univ. Paris-Sud, Univ. Paris-Saclay, 1 Avenue Augustin Fresnel, 91767 Palaiseau, France }
 \author{Ursula Ebels}\affiliation{ Univ. Grenoble Alpes, CEA, INAC-SPINTEC, CNRS, SPINTEC, 38000 Grenoble, France }
\author{Akio Fukushima}\affiliation{National Institute of Advanced Industrial Science and Technology (AIST),
Spintronics Research Center, Tsukuba, Ibaraki 305-8568, Japan}
\author{Hitoshi Kubota}\affiliation{National Institute of Advanced Industrial Science and Technology (AIST),
Spintronics Research Center, Tsukuba, Ibaraki 305-8568, Japan} 
 \author{Shinji Yuasa}\affiliation{National Institute of Advanced Industrial Science and Technology (AIST),
Spintronics Research Center, Tsukuba, Ibaraki 305-8568, Japan}
\author{Julie Grollier}\affiliation{Unit\'{e} Mixte de Physique CNRS/Thales, Univ. Paris-Sud, Univ. Paris-Saclay, 1 Avenue Augustin Fresnel, 91767 Palaiseau, France }
\author{Gilles Cibiel}\affiliation{Centre National d'\'{E}tudes Spatiales (CNES), 18 av. Edouard Belin, 31401 Toulouse, France}
\author{Serge Galliou}\affiliation{FEMTO-ST Institute, CNRS, Univ. Bourgogne Franche Comt\'{e}, 25030 Besançon, France}
\author{Enrico Rubiola}\affiliation{FEMTO-ST Institute, CNRS, Univ. Bourgogne Franche Comt\'{e}, 25030 Besançon, France}
 \author{Vincent Cros}\email[]{vincent.cros@cnrs-thales.fr}\affiliation{Unit\'{e} Mixte de Physique CNRS/Thales, Univ. Paris-Sud, Univ. Paris-Saclay, 1 Avenue Augustin Fresnel, 91767 Palaiseau, France }


\date{\today}

\definecolor{light-gray}{rgb}{0.96,0.96,0.96}

%
%
%
%
%
%
%
%

\pacs{}

\maketitle



\section*{Supplementary}

\subsection{Thiele simulation}

The simulation of the vortex dynamics is performed based on the differential Thiele equation \cite{Thiele1973}. It describes the spin transfer torque iduced gyrotropic motion of the vortex core well.
Following the description in Refs. \cite{Dussaux2012,Grimaldi2017}, we also consider higher order terms of damping and confinement, what allows a deterministic description of the dynamics through the normalized oscillation radius $s(t)$ and phase $\theta(t)$ of the vortex core in the nanodisk:
\begin{align*}
\dot{\theta} &= \frac{\kappa}{G} \left( 1+\zeta s^2 \right)  \\
\dot{s} &= \frac{D_0\kappa s}{G^2} \left( \frac{a_j I G }{D_0 \kappa \pi R^2} -  1  +  ( \zeta+\xi )s^2   \right)   ~~~.
\end{align*}

The different parameters are summarized in table \ref{tab:table1}, including their meaning and value chosen for the performed simulations.
Furthermore, $I$ is the applied dc current, and $\kappa(1+\zeta s^2)$ the confinement stiffness with $\kappa$ its linear part and $\zeta$ its nonlinearity factor. In the table, we show the values of the magnetostatic confinement $\kappa_{ms}$ with nonlinear part $\kappa_{ms}'$, and the Oersted field confinement $\kappa_{Oe}$ with nonlinear part $\kappa_{Oe}'$. It is $\kappa = \kappa_{ms} + \kappa_{Oe} I/(\pi R^2)$ and $\zeta = \frac{\kappa_{ms}' + \kappa_{Oe}' I/(\pi R^2)}{\kappa_{ms} + \kappa_{Oe} I/(\pi R^2)}$.

\begin{table}[bth!]
	\begin{ruledtabular}
		\begin{tabular}{p{0.56\columnwidth}p{0.414\columnwidth}}
			$R=187.5\,$nm & nano-dot radius\\
			$D_{0}= 4.28 \cdot 10^{-15}\,$kg.rad$^{-1}$.s$^{-1}$ & linear damping coefficient\\
			$\xi=0.6$ & nonlinear damping coefficient\\
			$G=2.0 \cdot 10^{-13}\,$kg.rad$^{-1}$.s$^{-1}$ & gyrovector amplitude\\
			$a_j=3.9\cdot 10^{-16}\,$kg.m$^{2}$.A$^{-1}$.s$^{-2}$ & spin-transfer torque efficiency\\
			$\kappa_{ms}=4.05 \cdot 10^{-4}\,$kg.s$^{-2}$ & magnetostatic confinement\\
			$\kappa_{ms}^{'}=1.01 \cdot 10^{-4}\,$kg.s$^{-2}$ & nonlinear magnetostatic confinement\\
			$\kappa_{Oe}=1.42 \cdot 10^{-15}\,$kg.m$^{2}$.A$^{-1}$.s$^{-2}$ & Oersted field confinement\\
			$\kappa_{Oe}^{'}=-7.12 \cdot 10^{-16}\,$kg.m$^{2}$.A$^{-1}$.s$^{-2}$ & nonlinear Oersted field confinement\\
		\end{tabular}
	\end{ruledtabular}
	\caption{\label{tab:table1} Parameters used for the simulation of the vortex dynamics in presence of thermal and flicker noise.}
\end{table}

\subsubsection*{Flicker noise generation}

In order to introduce the flicker noise in the Thiele equation framework, we model a generating noise process of $1/f^1$ spectral shape, as shown in fig. 1 of the main text. This is then converted into the characteristic amplitude and phase noise PSD due to the dynamical nonlinear stochastic Langevin differential equations \cite{Wittrock2019_PRB}.
We define  a random variable $r_{1/f}(t)$ that has a $1/f^1$ flicker noise PSD and is scaled by a factor $\lambda$. This random process is then added to the applied dc current $I_{dc}$:
\begin{equation}
I_{STO}(t)=  I_{dc} + \lambda r_{1/f}(t)   ~~~.
\label{eq:simulation_ISTO_flicker}
\end{equation}
As elaborated in the main text, $\lambda=2.5\cdot 10^{-4}$ was chosen in such a way that the final amplitude and phase noise PSD are meaningful and similar to those observed experimentally. 
In order to produce the random stream variable $r_{1/f}(t)$, we choose as a first step to construct a Fourier series decomposition with Fourier amplitude coefficients that have a $1/f$ decay and uniformly distributed random phase coefficients. By taking the inverse Fourier transform of this constructed decomposition, we obtain our $1/f$ random stream variable $r_{1/f}(t)$. As shown in fig. 1 of the main text, the noise PSD of the injected dc current $I_{STO}(t)$ in the simulation follows a $1/f$ relation for a broadband frequency bandwidth.

\subsection{Discussion of the generating noise}

The flicker generating noise, which leads to the $1/f^{\beta}$ power law characteristics in the amplitude and phase noise PSDs of the STVO, itself exhibits a $1/f^1$ spectral shape \cite{Wittrock2019_PRB}. Its fundamental origin is not yet well understood, and in the main paper we model this original generating noise to be caused by the supplying dc source. 

\begin{figure}[bth!]
		\centering
		\vspace{1em}
		\includegraphics[width=0.74\columnwidth]{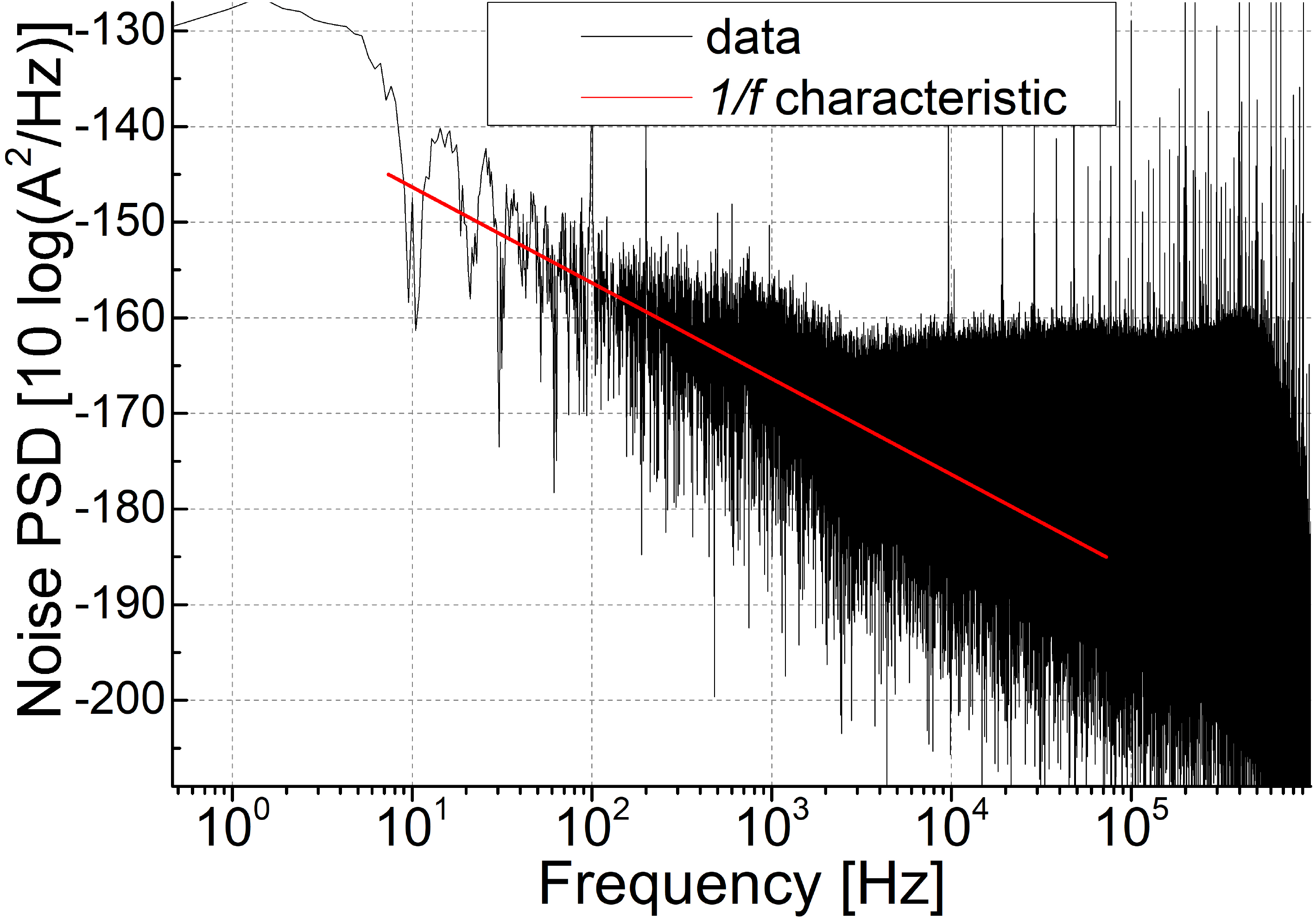}
		\caption[]{$1/f$ flicker noise characteristics of the supplying Keithley 6220 dc current source.}
		\label{fig_exp:generating-noise_current-source}
		\end{figure}
To gain more insight into the particular origins of the flicker noise and the influence of the current source characteristics, we can compare the noise PSD of the modelled generating noise (in the main paper) with the experimental noise PSD of the current source.
In fig. \ref{fig_exp:generating-noise_current-source}, we present the measured noise characteristics of the used Keithley 6220 current source on the circuit with a $50\,\Omega$ resistance. 		
The shown data are measured at $8\,$mA. Indeed, the noise PSD does not change much within the current interval $]0;10]\,$mA. Above, the current source changes range and the noise PSD might be different.		

We see that indeed the noise PSD exhibits a $1/f$ spectral shape.
Comparing the modelled generating noise with the measured noise PSD of the current supply, we find that the magnitude of the modelled noise is much higher than the one from the measurement, i.e. for instance at $10^4\,$Hz $\sim -155\,$dB in the simulation compared to  $\sim -180\,$dB for the measurement.
We can conclude here that indeed different mechanisms contribute to the characteristic $1/f$ flicker noise. These might be fluctuations in the magnetic texture, the impact of defects and/or inhomogeneities in the magnetic layers or the tunnel barrier, or also external fluctuations of the applied magnetic field. Hence, also the noise of the dc current supply  contributes but is not the major origin for the $1/f$ flicker noise in spin torque oscillators.
However, in the simulation we model all these different origins of $1/f$ noise to be included in the dc current fluctuations.

%
%
%



\subsection{Phase variance $\leftrightarrow$ phase noise relation}

Taking the Wiener-Khintchine theorem, the autocorrelation of the phase fluctuations $\delta \phi$ can be expressed through the phase noise PSD. Taking also the definition of the autocorrelation, we get:
\begin{align*}
K_{\delta \phi} (\tau) = \frac{1}{2\pi} \int\limits_{-\infty}^{\infty} S_{\delta \phi} (\omega) e^{i\omega \tau} d\omega = \braket{ \delta \phi (t+\tau)~\delta \phi(t) } 
\end{align*}

We define the phase difference at times $t$ and $\tau$:
\begin{align*}
\psi(t,\tau) = \phi(t+\tau) -\phi(t) = \delta\phi(t+\tau) - \delta\phi(t)
\end{align*}

Its variance function is:
\begin{align*}
\Delta \phi ^2  &= \braket{\psi^2} - \braket{\psi}^2 = \braket{ \left[\delta \phi (t+\tau) - \delta \phi(t)   \right]^2 } - 0 \\
&= \braket{\delta\phi(t+\tau)^2} + \braket{\delta\phi(t)^2} - 2\braket{\delta \phi (t+\tau)\delta\phi(t)} \\
&= 2 \left[  \braket{ \delta\phi^2 } - \braket{\delta\phi(t+\tau)\delta\phi(t)}     \right]
\end{align*}

This means that the variance function $\Delta\phi^2$ can be expressed in terms of phase noise:
\begin{align*}
\frac{\Delta\phi^2}{2} &= K_{\delta\phi}(0) - K_{\delta\phi}(\tau) = \frac{1}{2\pi} \int\limits_{\mathbb{R}} \left( 1 - e^{i \omega t }  \right) \cdot S_{\delta \phi }(\omega)  d\omega \\ 
&= \frac{1}{2\pi} \int\limits_{\mathbb{R}} \left( 1 - \cos( \omega t) - i \sin(\omega t)  \right) \cdot S_{\delta \phi }  d\omega  
\end{align*}
\noindent
Because $S_{\delta \phi }(\omega)$, similar to the autocorrelation function $K(t)$, is always an even function (Cauchy principal value), the expression can be simplified to:
\begin{align}
\frac{\Delta\phi^2}{2}  &=   \frac{1}{\pi} \int\limits_{0}^{\infty} \left( 1 - \cos( \omega t) \right) \cdot S_{\delta \phi }  d\omega      ~~~. \label{eq:noise-PSD_2}
\end{align}

\subsection{Phase variance $\leftrightarrow$ autocorrelation relation}

Here, we discuss the autocorrelation function of the oscillation signal $c=\sqrt{p}e^{i\phi}$, not of the fluctuations as done above.
The signal power is $p=p_0+\delta p$ and the phase $\phi(t) = -\omega(p_0)t+\phi_i + \delta\phi(t)$. 
We assume the power fluctuations little compared to the phase fluctuations: $\delta p \ll \delta \phi$.
It is for the autocorrelation:
\begin{align*}
K(\tau) &= \braket{\sqrt{p_0+\delta p} ~ e^{i\phi(t+\tau)} \sqrt{p_0 + \delta p} ~ e^{-i\phi(t)} } \\ &\approx  p_0 \braket{e^{i\left[ \phi(t+\tau) - \phi(t)       \right] } } 
\approx p_0 e^{-i\omega(p_0)\tau} \braket{e^{i\left[ \delta\phi(t+\tau) - \delta \phi(t)       \right] } } \\
& = p_0 e^{-i\omega(p_0)\tau} \braket{e^{i\psi (t,\tau) } }
\end{align*}

As defined already above, we used $\psi (t,\tau) = \delta\phi(t+\tau) - \delta\phi(t)$.

Because the fluctuations are defined by a stationary ergodic process following a 0-centered Gaussian law, $\psi$ has the following probability function:
\begin{align*}
p(\psi) = \frac{1}{\Delta \phi \sqrt{2\pi}} e^{-\frac{\psi^2}{2\Delta \phi^2}}   ~~~.
\end{align*}

Using this, it follows:
\begin{gather*}
\braket{e^{i\psi (t,\tau) } } = \int\limits_{-\infty}^{\infty} e^{i\psi} p(\psi) d\psi \\
=  \int\limits_{-\infty}^{\infty} \left[ \cos(\psi)+\underbrace{i\sin(\psi)}_{\text{antisymmetric}\rightarrow =0} \right] p(\psi) d\psi \\
= \frac{1}{\Delta\phi \sqrt{2\pi}} \int\limits_{-\infty}^{\infty} \cos(\psi) e^{-\psi^2 / (2\Delta\phi^2)}  d\psi 
= e^{-\Delta\phi^2 /2} ~~~.
\end{gather*}
In the last step, the Gaussian integral was evaluated.

In total, we get the final result:
\begin{align*}
K(\tau) \approx  p_0 e^{-i\omega(p_0)\tau} e^{-\Delta\phi^2/2 }    ~~~.
\end{align*}

\bibliography{literatur_promo}